\documentclass[twocolumn]{aastex61}

\newcommand\hst{{\em HST}}
\newcommand\swift{{\em Swift}}

\newcommand{\SuperNu}{{\tt SuperNu}}
\newcommand{\WinNet}{{\tt WinNet}}

\newcommand{\sss}{AT2017gfo}
\newcommand{\KN}{KN/MN}

\newcommand{\HST}{\textit{HST}}

\newcommand{\Ks}{\ensuremath{K_{\rm s}}}

\submitjournal{ApJL}

\shorttitle{The emergence of a kilonova following the merger of two neutron stars}
\shortauthors{Tanvir et al.}

\begin{document}

\title{The Emergence of a Lanthanide-Rich Kilonova Following the Merger of Two Neutron Stars}

\correspondingauthor{N. R. Tanvir}
\email{nrt3@le.ac.uk}

\author{N. R. Tanvir}
\affiliation{University of Leicester, Department of Physics \& Astronomy and Leicester Institute of Space \& Earth Observation,  University Road,  Leicester, LE1 7RH, United Kingdom}
\author{A. J. Levan}
\affiliation{Department of Physics, University of Warwick, Coventry, CV4 7AL, United Kingdom}
\author{C. Gonz\'alez-Fern\'andez}
\affiliation{Institute of Astronomy, University of Cambridge, Madingley Road, Cambridge, CB3 0HA, United Kingdom}

\author{O. Korobkin}
\affiliation
{Computational Methods Group (CCS-2), Los Alamos National Laboratory, P.O. Box 1663, Los Alamos, NM, 87545, USA}

\author{I. Mandel}
\affiliation{Birmingham Institute for Gravitational Wave Astronomy and School of Physics and Astronomy, University of Birmingham, Birmingham, B15 2TT, UK}
\author{S. Rosswog}
\affiliation{The Oskar Klein Centre, Department of Astronomy, AlbaNova, Stockholm University, SE-106 91 Stockholm, Sweden}
\author{J. Hjorth}
\affiliation{Dark Cosmology Centre, Niels Bohr Institute, University of Copenhagen, Juliane Maries Vej 30, 2100 Copenhagen \O,  Denmark}


\author{P. D'Avanzo}
\affiliation{INAF, Osservatorio Astronomico di Brera, Via E. Bianchi 46, I-23807 Merate (LC), Italy}

\author{A. S. Fruchter}
\affiliation{Space Telescope Science Institute, 3700 San Martin Drive, Baltimore, MD 21218, USA}

\author{C. L. Fryer}
\affiliation
{Computational Methods Group (CCS-2), Los Alamos National Laboratory, P.O. Box 1663, Los Alamos, NM, 87545, USA}

\author{T. Kangas}
\affiliation{Space Telescope Science Institute, 3700 San Martin Drive, Baltimore, MD 21218, USA}

\author{B. Milvang-Jensen}
\affiliation{Dark Cosmology Centre, Niels Bohr Institute, University of Copenhagen, Juliane Maries Vej 30, 2100 Copenhagen \O, Denmark}

\author{S. Rosetti}
\affiliation{University of Leicester, Department of Physics \& Astronomy and Leicester Institute of Space \& Earth Observation,  University Road,  Leicester, LE1 7RH, United Kingdom}

\author{D. Steeghs}
\affiliation{Department of Physics, University of Warwick, Coventry, CV4 7AL, United Kingdom}

\author{R. T. Wollaeger}
\affiliation
{Computational Methods Group (CCS-2), Los Alamos National Laboratory, P.O. Box 1663, Los Alamos, NM, 87545, USA}

\author{Z. Cano}
\affiliation{Instituto de Astrof\'isica de Andaluc\'ia (IAA-CSIC), Glorieta de la Astronom\'ia s/n, 18008 Granada, Spain}

\author{C. M. Copperwheat}
\affiliation{Astrophysics Research Institute, Liverpool John Moores University, Liverpool Science Park IC2, 146 Brownlow Hill, Liverpool L3 5RF, UK}

\author{S. Covino}
\affiliation{INAF, Osservatorio Astronomico di Brera, Via E. Bianchi 46, I-23807 Merate (LC), Italy}

\author{V. D'Elia}
\affiliation{Space Science Data Center, ASI, Via del Politecnico, s.n.c., 00133, Roma, Italy}
\affiliation{INAF, Osservatorio Astronomico di Roma, Via di Frascati, 33, I-00078 Monteporzio Catone, Italy}

\author{A. de Ugarte Postigo}
\affiliation{Instituto de Astrof\'isica de Andaluc\'ia (IAA-CSIC), Glorieta de la Astronom\'ia s/n, 18008 Granada, Spain}
\affiliation{Dark Cosmology Centre, Niels Bohr Institute, University of Copenhagen, Juliane Maries Vej 30, 2100 Copenhagen \O,  Denmark}

\author{P. A. Evans}
\affiliation{University of Leicester, Department of Physics \& Astronomy and Leicester Institute of Space \& Earth Observation,  University Road,  Leicester, LE1 7RH, United Kingdom}

\author{W. P. Even}
\affiliation
{Computational Methods Group (CCS-2), Los Alamos National Laboratory, P.O. Box 1663, Los Alamos, NM, 87545, USA}

\author{S. Fairhurst}
\affiliation{School of Physics and Astronomy, Cardiff University, Cardiff, UK}

\author{R. Figuera Jaimes}
\affiliation{SUPA, School of Physics \& Astronomy, University of St Andrews, North Haugh, St Andrews KY16 9SS, UK}

\author{C. J. Fontes}
\affiliation
{Computational Methods Group (CCS-2), Los Alamos National Laboratory, P.O. Box 1663, Los Alamos, NM, 87545, USA}

\author{Y. I. Fujii}
\affiliation
{Niels Bohr Institute \& Centre for Star and Planet Formation, University of Copenhagen {\O}ster Voldgade 5, 1350 - Copenhagen, Denmark}
\affiliation
{Institute for Advanced Research, Nagoya University, Furo-cho, Chikusa-ku, Nagoya, 464-8601, Japan}

\author{J. P. U. Fynbo}
\affiliation{Niels Bohr Institute, University of Copenhagen, Juliane Maries Vej 30, 2100 Copenhagen \O,  Denmark}

\author{B. P. Gompertz}
\affiliation{Department of Physics, University of Warwick, Coventry, CV4 7AL, United Kingdom}

\author{J. Greiner}
\affiliation{Max-Planck-Institut f\"ur extraterrestrische Physik, 85740 Garching,
    Giessenbachstr. 1, Germany}

\author{G. Hodosan}
\affiliation{Instituto de Astrof\'isica de Andaluc\'ia (IAA-CSIC), Glorieta de la Astronom\'ia s/n, 18008 Granada, Spain}

\author{M. J. Irwin}
\affiliation{Institute of Astronomy, University of Cambridge, Madingley Road, Cambridge, CB3 0HA, United Kingdom}

\author{P. Jakobsson}
\affiliation{Centre for Astrophysics and Cosmology, Science Institute, University of Iceland, Dunhagi 5, 
107 Reykjav\'ik, Iceland}

\author{U. G. J{\o}rgensen}
\affiliation{Niels Bohr Institute \& Centre for Star and Planet Formation, University of Copenhagen {\O}ster Voldgade 5, 1350 - Copenhagen, Denmark
}

\author{D. A. Kann}
\affiliation{Instituto de Astrof\'isica de Andaluc\'ia (IAA-CSIC), Glorieta de la Astronom\'ia s/n, 18008 Granada, Spain}

\author{J. D. Lyman}
\affiliation{Department of Physics, University of Warwick, Coventry, CV4 7AL, United Kingdom}

\author{D. Malesani}
\affiliation{Niels Bohr Institute, University of Copenhagen, Juliane Maries Vej 30, 2100 Copenhagen \O,  Denmark}

\author{R. G. McMahon}
\affiliation{Institute of Astronomy, University of Cambridge, Madingley Road, Cambridge, CB3 0HA, United Kingdom}

\author{A. Melandri}
\affiliation{INAF, Osservatorio Astronomico di Brera, Via E. Bianchi 46, I-23807 Merate (LC), Italy}

\author{P.T. O'Brien}
\affiliation{University of Leicester, Department of Physics \& Astronomy and Leicester Institute of Space \& Earth Observation,  University Road,  Leicester, LE1 7RH, United Kingdom}

\author{J. P. Osborne}
\affiliation{University of Leicester, Department of Physics \& Astronomy and Leicester Institute of Space \& Earth Observation,  University Road,  Leicester, LE1 7RH, United Kingdom}

\author{E. Palazzi}
\affiliation{INAF, Istituto di Astrofisica Spaziale e Fisica Cosmica, Via Gobetti 101, I-40129 Bologna, Italy}

\author{D. A. Perley}
\affiliation{Astrophysics Research Institute, Liverpool John Moores University, Liverpool Science Park IC2, 146 Brownlow Hill, Liverpool L3 5RF, UK}

\author{E. Pian}
\affiliation{INAF, Istituto di Astrofisica Spaziale e Fisica Cosmica, Via Gobetti 101, I-40129 Bologna, Italy}

\author{S. Piranomonte}
\affiliation{INAF, Osservatorio Astronomico di Roma, Via di Frascati, 33, I-00078 Monteporzio Catone, Italy}

\author{M. Rabus}
\affiliation{Instituto de Astrof\'isica, Pontificia Universidad Cat\'olica de Chile, Av. Vicu\~na Mackenna 4860, 7820436 Macul, Santiago, Chile}

\author{E. Rol}
\affiliation{School of Physics and Astronomy, Monash University, PO Box
27, Clayton, Victoria 3800, Australia}

\author{A. Rowlinson}
\affiliation{Anton Pannekoek Institute, University of Amsterdam, Science Park 904, 1098 XH Amsterdam, the Netherlands}
\affiliation{ASTRON, the Netherlands Institute for Radio Astronomy, Postbus 2, 7990 AA Dwingeloo, the Netherlands}

\author{S. Schulze}
\affiliation{Department of Particle Physics and Astrophysics, Weizmann Institute of Science, Rehovot 761000, Israel}

\author{P. Sutton}
\affiliation{School of Physics and Astronomy, Cardiff University, Cardiff, UK}

\author{C.C. Th\"one}
\affiliation{Instituto de Astrof\'isica de Andaluc\'ia (IAA-CSIC), Glorieta de la Astronom\'ia s/n, 18008 Granada, Spain}

\author{K. Ulaczyk}
\affiliation{Department of Physics, University of Warwick, Coventry, CV4 7AL, United Kingdom}

\author{D. Watson}
\affiliation{Niels Bohr Institute, University of Copenhagen, Juliane Maries Vej 30, 2100 Copenhagen \O,  Denmark}

\author{K. Wiersema}
\affiliation{University of Leicester, Department of Physics \& Astronomy and Leicester Institute of Space \& Earth Observation,  University Road,  Leicester, LE1 7RH, United Kingdom}

\author{R.A.M.J. Wijers}
\affiliation{Anton Pannekoek Institute, University of Amsterdam, Science Park 904, 1098 XH Amsterdam, the Netherlands}




\begin{abstract}
We report the discovery and monitoring of the near-infrared counterpart (\sss) of a binary neutron-star merger event
detected as a gravitational wave source by Advanced LIGO/Virgo (GW170817) and as a short gamma-ray burst by \emph{Fermi}/GBM and \emph{Integral}/SPI-ACS (GRB\,170817A). The evolution of the transient light is consistent with predictions for the behaviour of a ``kilonova/macronova", powered by the radioactive decay of massive neutron-rich nuclides created via r-process nucleosynthesis in the neutron-star ejecta.
In particular, evidence for this scenario is found from broad features seen in {\em Hubble Space Telescope} infrared spectroscopy, similar to those predicted for lanthanide dominated ejecta, and the much slower evolution in the near-infrared \Ks-band compared to the optical. This indicates that the late-time light is dominated by high-opacity lanthanide-rich ejecta, suggesting nucleosynthesis to the 3rd r-process peak (atomic masses $A\approx195$).
This discovery confirms that neutron-star mergers produce kilo-/macronovae 
and that they are at least a major -- if not the dominant -- site of 
rapid neutron capture nucleosynthesis in the universe.
\end{abstract}




\section{Introduction} \label{sec:intro}

When compact binary star systems merge, they release copious amounts of energy in the form of gravitational waves \citep[GWs][]{abbott16,abbott17}. 
If the system is either a binary neutron star (BNS) or a neutron star and stellar mass black hole (NSBH), the merger is expected to be accompanied by various electromagnetic phenomena. In particular, systems of this sort have long been thought to be the  progenitors of short-duration gamma-ray bursts 
\citep[short-GRBs; e.g.][]{eichler89, nakar07}, 
whilst their neutron-rich ejecta should give rise to a so-called ``kilonova" or ``macronova" (\KN) explosion \citep{li98,kulkarni05,rosswog05,metzger10}. 
Short-GRBs are bright and conspicuous high-energy events. However, since they are thought to be jetted systems, they are 
expected to be observed for only a subset of such mergers, as the most intense emission from a given merger will usually not intersect our line of sight. \KN, which are powered by radioactive decay, although considerably fainter,
emit more isotropically \citep[e.g.][]{grossman14} and peak later than short-GRB afterglows.  Thus, they 
 are generally considered to provide the best prospects for electromagnetic
(EM) counterparts to GW detections
\citep[e.g.][]{metzger12,kelley13,fernandez16,rosswog17}.

However, it has been argued that the high opacity of newly synthesised heavy elements in the \KN\ ejecta, particularly lanthanides and actinides, will render them faint in the optical, 
with emission instead appearing primarily in the near-infrared on timescales of several days \citep{kasen13,barnes13,tanaka13}.
This connects them closely to cosmic nucleosynthesis.
The ``rapid neutron capture" or ``r-process"
is responsible for about half of the elements heavier than iron and had
traditionally been attributed to core collapse supernovae \citep{burbidge57}. 
A number of recent studies, however,  have disfavored supernovae since their 
conditions were found unsuitable for producing at least the heaviest elements 
of the ``platinum peak" near atomic mass $A= 195$. At the same time, neutron star mergers 
have gained increasing attention
as a major r-process production site. 
\cite{lattimer74} first discussed such compact binary mergers as an r-process 
site and since the first nucleosynthesis calculations 
\citep{rosswog98,freiburghaus99} a slew of other studies 
\citep[e.g.][]{goriely11,korobkin12,just15,mendoza_temis15} have
confirmed their suitability for the production of the heaviest elements in the  Universe.

To date, the most compelling evidence in support of this scenario was provided by the observation of excess infrared light (rest frame $\lambda\sim1.2$\,$\mu$m) at the location of a short-GRB about a week (in the rest frame) after the burst occurred \citep[GRB\,130603B;][]{tanvir13,berger13}. Subsequent work has uncovered possible ``kilonova" components in several other short-GRBs \citep{yang15,jin15,jin16}, although other late-time emission processes cannot be ruled out, and the fact that in these instances the excess was in the rest frame optical bands suggested it did not originate in lanthanide-rich ejecta.

While the initial focus has been on the extremely neutron-rich, low electron fraction ($Y_e$), ``tidal" 
ejecta component, recent studies \citep{perego14,wanajo14,just15,radice16} 
have highlighted that this material  is likely complemented by higher 
$Y_e$ material that still undergoes r-process nucleosynthesis, but does 
not produce the heaviest elements (such as gold or platinum) in the 
3rd r-process peak. This higher $Y_e$-material results from either shocks,
neutrino-driven winds and/or the unbinding of  the
accretion torus that is formed in the merger.  Being free of lanthanides, 
this material possesses lower opacities and produces earlier and bluer
optical transients \citep[e.g.][]{metzger14,kasen15}. Geometrically, the
low-$Y_e$, high opacity matter is ejected preferentially in the binary orbital
plane, while the higher-$Y_e$, low opacity ejecta is concentrated
towards the binary rotation axis. Existing numerical studies suggest that
dynamical ejecta has higher velocities
\citep[$>0.1c$; e.g.][]{kasen15,rosswog17}
and could -- if viewed edge-on -- 
obscure the wind-type ejecta. Therefore, significant viewing-angle effects 
are expected for the EM signatures of neutron-star mergers.

Here we present the optical and infrared light curve of an explosive transient seen in the hours and days following the detection of a BNS merger by 
Advanced LIGO/Virgo. 
We also present optical and near-infrared spectra of the transient.
The data show a marked colour change from blue to red on a time-scale of days as
well as conspicuous spectral features, 
strongly indicative of a kilonova showing both rapidly evolving blue and more slowly evolving red components. 

We use AB magnitudes throughout and, 
except where otherwise stated,
correct for Milky Way foreground extinction
according to $A_{\rm V}=0.338$ mag from \citet{schlafly11}.

\section{Observations}
\label{sec:obs}

The discovery of GW\,170817 by LIGO and Virgo was announced to electromagnetic follow-up partners shortly after the trigger time of 12:41:04 UT on 17 Aug 2017 \citep{GCN21509}. 
The potential importance of this event was immediately realised due to its temporal and (within the large error bounds) spatial coincidence with a short-duration GRB (170817A)
detected by {\em Fermi}/GBM at 12:41:06.47 UT
\citep{goldstein17}
and also {\em INTEGRAL}/SPI-ACS \citep{GCN21507,savchenko17}.
The existence of a short gamma-ray signal could be interpreted as requiring a close to pole-on viewing angle, but the absence of a normal 
GRB afterglow in subsequent monitoring \citep[e.g. in X-rays;][]{evans17}
instead suggests the possibility of
some kind of off-axis emission mechanism, such as may be produced by a shocked cocoon around the primary jet \citep[e.g.][]{lazzati17,gottlieb17}.

\subsection{Imaging}

\begin{figure*}[ht!]
\plotone{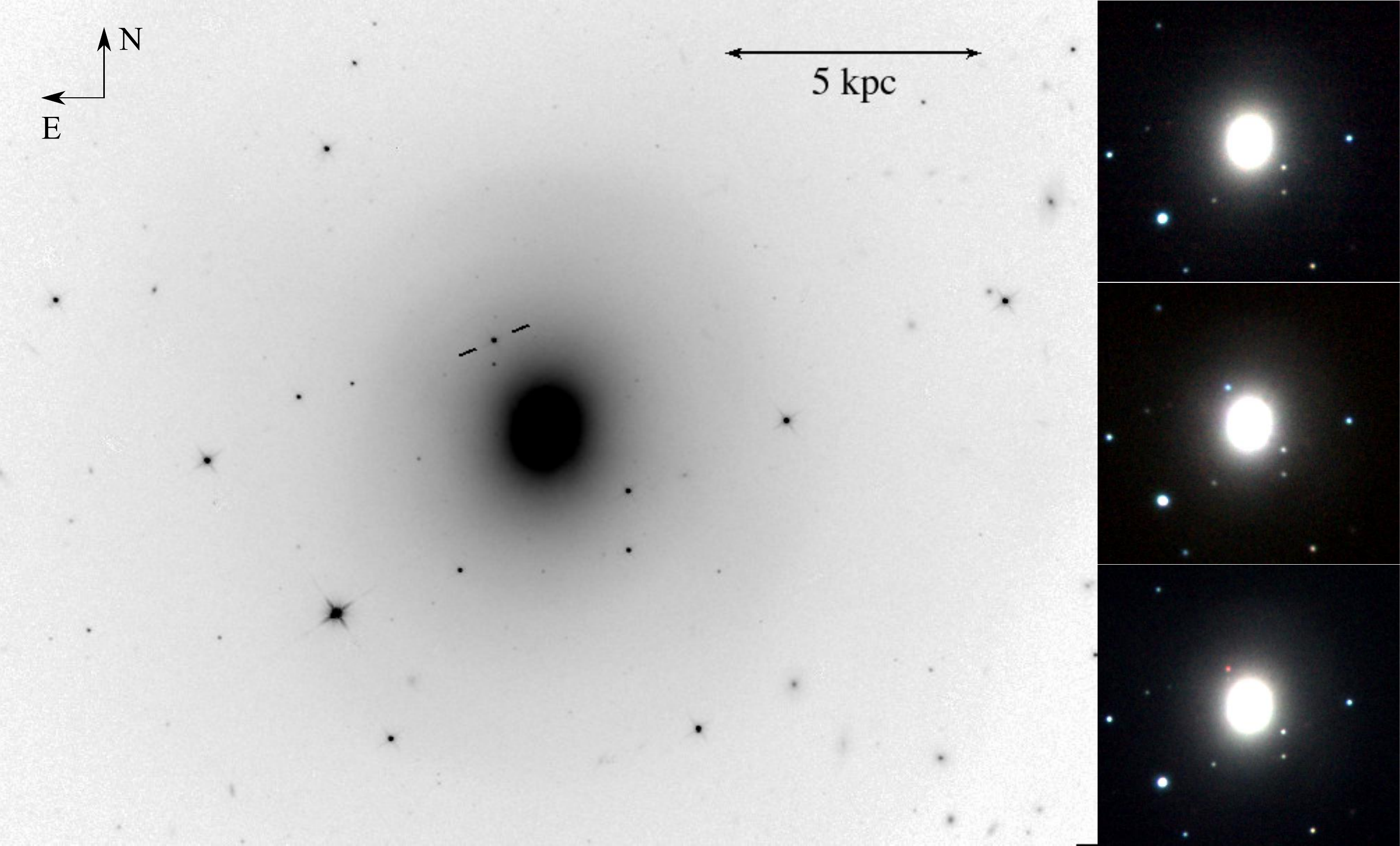}
\caption{Main panel shows the first epoch F110W \hst/WFC3-IR image of the field of \sss\ indicating its location within NGC\,4993.
The physical scale assuming a distance of 40\,Mpc is shown.
The sequence of panels on the right show
VISTA imaging (RGB rendition created
from $Y, J, K_s$ images) from pre-discovery (2014; top), discovery (middle) and at 8.5 days post-merger as the transient was fading and becoming increasingly red (bottom).
\label{fig:field}}
\end{figure*}

We triggered observations with the European Southern Observatory (ESO) Visible and Infrared Survey Telescope for Astronomy 
\citep[VISTA;][]{sutherland15} covering two fields within  the GW error region and containing high densities of galaxies in the plausible distance regime to have produced such a signal \citep{GCN21513}. 
Observations began in Chilean twilight at 23:24 UT using the $Y$ (1.02\,\micron), $J$ (1.25\,\micron) and \Ks\ (2.15\,\micron) filters. In the second field we identified a bright new point source, visible in all three filters, which was not apparent in prior imaging of the field obtained as part of the VISTA Hemisphere Survey \citep{mcmahon13}. These images were processed using a tailored version of the VISTA Data Flow System that follows the standard reduction path described in \citet{cgf17} but allows for quick processing of data by using the most current set of calibration frames (mainly flat fields) available at the time. 
The sky location of the transient was 
$\textnormal{RA}(2000)=13$:09:48.09, $\textnormal{Dec.}(2000)=-$23:22:53.3, approximately 10$^{\prime\prime}$ from the centre of the S0 galaxy NGC\,4993 (Figure~\ref{fig:field}). 
Contemporaneous observations made independently with several optical telescopes also revealed 
a new source at this location \citep{coulter17,GCN21530,valenti17}, which was designated \sss\ (also referred to as SSS17a and DLT17ck).

Subsequently we monitored \sss\ with VISTA at roughly nightly cadence until the field became too difficult to observe due to its proximity to the Sun, 
after $\sim25$ days. At later epochs, observations were restricted to the \Ks-band, which is least affected by twilight observing.

Additionally, we imaged the field with the ESO Very Large Telescope (VLT), the {\em Hubble Space Telescope} ({\em HST}),
the Nordic Optical Telescope (NOT), and the Danish 1.5\,m Telescope (DK1.5), including optical observations (a full list of observations and description of photometric measurements is given in Table~\ref{tab:oir}). VLT observations were taken with 
VIMOS and HAWK-I in the optical ($r$, $z$), and infrared ($K$) bands respectively. Observations were processed through {\tt esorex} in a standard fashion. {\em HST} observations were obtained in the optical (F475W, F606W and F814W) and IR (F110W \& F160W) reduced using {\tt astrodrizzle} to combine, distortion correct and cosmic-ray reject individual images. The images were ultimately drizzled to plate scales of 0.025\arcsec pixel$^{-1}$ (for UVIS) and 0.07\arcsec pixel$^{-1}$ (for the IR). 

For each  image, the light from the host galaxy was modeled and subtracted using custom routines, to aid
photometry of the transient, which was performed using the {\tt GAIA} software\footnote{\url{http://star-www.dur.ac.uk/~pdraper/gaia/gaia.html}}.
The ground-based $J$- and \Ks-bands were calibrated to the 2MASS\footnote{\url{http://irsa.ipac.caltech.edu/Missions/2mass.html}}
stars in the field, while the $Y$-band was calibrated via the relations given in \citet{cgf17}.
The optical filters
were calibrated to the Pan-STARRS\footnote{\url{http://archive.stsci.edu/panstarrs/}} scale.
The \HST\ photometry used the standard WFC3 calibrations\footnote{\url{http://www.stsci.edu/hst/wfc3/phot_zp_lbn}}, apart from the F110W observations which were also calibrated to the $J$-band to aid comparison with the other $J$-band photometry.

\begin{figure*}[ht!]
\begin{center}
\includegraphics[width=16cm,angle=0]{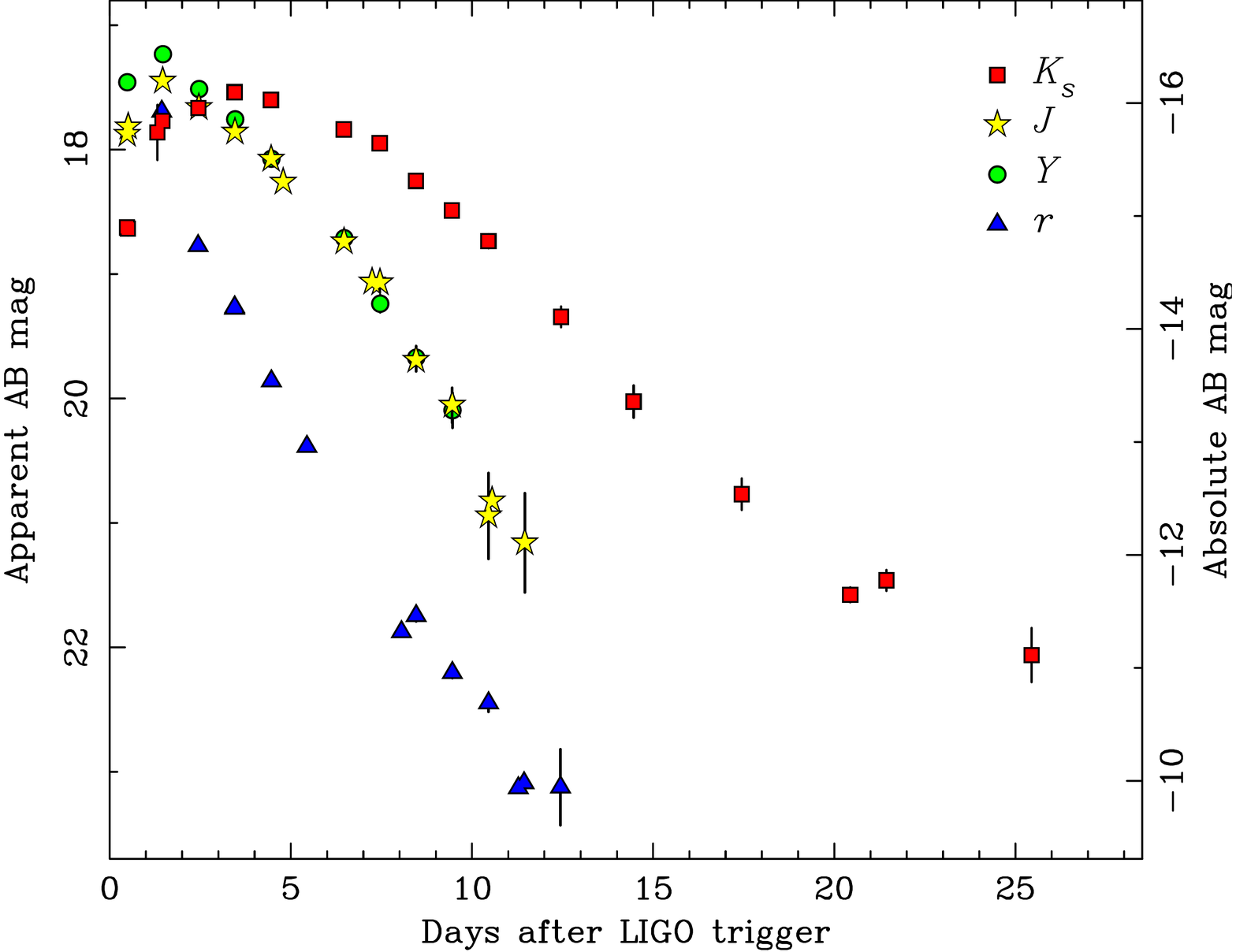}
\end{center}
\caption{The light curves of \sss\ in the $r$-, $Y$-, $J$- and \Ks-bands.
The absolute magnitude, assuming a distance of 40\,Mpc, is shown on the right hand scale. Note that in many cases the error bars are smaller than the symbols.
\label{fig:lc}}
\end{figure*}

\begin{figure}[ht!]
\hspace{-1mm}\includegraphics[width=8cm,angle=0]{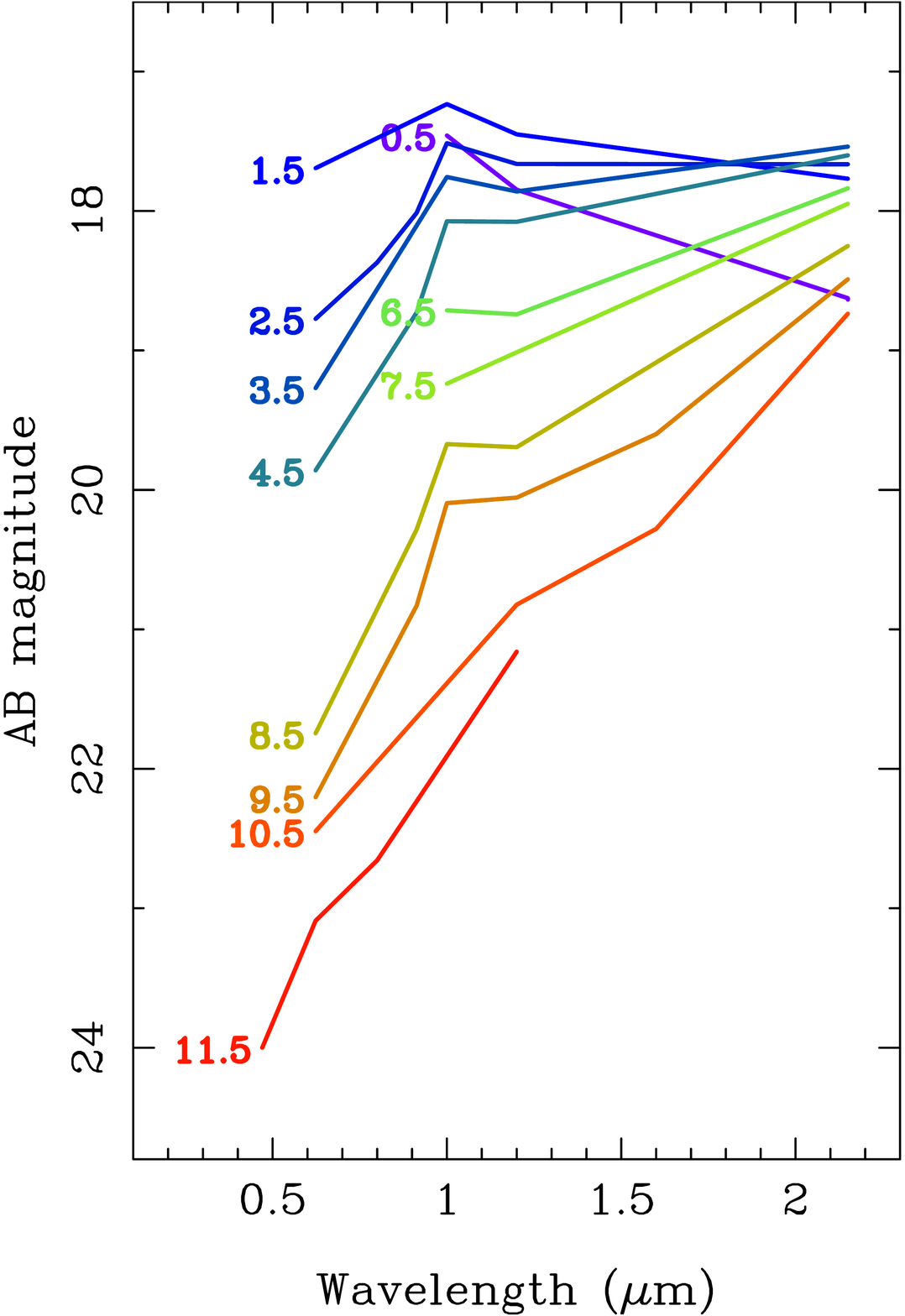}
\caption{The evolution of the broad-band spectral energy distribution 
of \sss\ over the first $\sim12$\,days
illustrating the marked blue to red trend.
\label{fig:sed}}
\end{figure}

Over the first several days \sss\ exhibited marked colour evolution from blue to red (Figures~\ref{fig:lc} \& \ref{fig:sed}).  Following a slow rise within the first day or so, the optical light declined rapidly from a peak in the first 36\,hr, and proceeded to follow an approximately exponential decline
(half-life in $r$-band $\approx40$\,hr). 
The $Y$- and $J$-band light curves track each other closely, and again decline following a peak in the first $\sim36$\,hr. By contrast, the  \Ks-band, 
exhibits a 
much broader peak than the optical,
varying by only $\approx20$\% in flux
from about 30\,hr to 6\,d post-merger.  

Although there is some evidence for dust lanes in the galaxy, its early-type nature and the absence
of host absorption lines (Section~\ref{sec:spec}) suggests little dust extinction.
Furthermore, the transient is located away from these obviously dusty regions (see \cite{levan17}. for details of host morphology and transient location).
This is supported by the linear polarimetry of the transient, which shows very low levels of polarisation \citep{covino17}, implying a line of sight dust column in the host galaxy of $E(B-V) \lesssim 0.2$\,mag (assuming a Milky Way 
like relation between $E(B-V)$ and linear polarisation).
Thus we only correct the photometry for dust extinction in the Milky Way.
The measured peak apparent magnitudes are $Y_{0}=17.22$ and $K_{0}=17.54$. 

\startlongtable
\begin{deluxetable}{cClcC}
\tablecaption{Optical and near-IR photometry of \sss \label{tab:oir}}
\tablecolumns{5}
\tablenum{1}
\tablewidth{0pt}
\tablehead{
\colhead{$\Delta t$ (d)} & \colhead{$t_{\rm exp}$ (s)} & \colhead{Telescope/Camera} & \colhead{Filter} & \colhead{Mag(AB)$_{\rm 0}$} }
\decimalcolnumbers
\startdata
8.116  &  520 & HST/WFC3-UVIS & F475W &   23.14\pm0.02\\
11.300  & 520  & HST/WFC3-UVIS & F475W &  24.08\pm0.05\\
11.411  & 600  & HST/WFC3-UVIS & F475W&  23.96\pm0.05\\
  1.44 & 30 & VLT/FORS & $r$& 17.69\pm0.02\\
  2.44 & 10 & VLT/FORS & $r$& 18.77\pm0.04\\
  3.45 & 60 & VLT/FORS & $r$& 19.28\pm0.01\\
  4.46 & 240 & VLT/VIMOS &$r$ & 19.86\pm0.01\\
  5.44 & 20 & VLT/FORS & $r$& 20.39\pm0.03\\
  8.46 & 600 & VLT/VIMOS &$r$ & 21.75\pm0.05\\
  9.46 & 600 & VLT/VIMOS &$r$ & 22.20\pm0.04\\
 10.46 &1200  & VLT/VIMOS &$r$ & 22.45\pm0.07\\
 11.44 & 360 & HST/WFC3-UVIS & F606W & 23.09\pm0.03\\
 12.44 &1200  & VLT/VIMOS &$r$ & 23.12\pm0.31\\
2.459   & 150 & DK1.5 & $i$&     18.37\pm0.03\\
11.428 & 560 & HST/WFC3-UVIS & F814W &   22.32\pm0.02\\
2.461  & 150 & DK1.5 & $z$&   18.01\pm0.13\\
4.451  & 240 & VLT/VIMOS & $z$&   18.73\pm0.01\\
8.443  & 400 & VLT/VIMOS & $z$&   20.28\pm0.03\\
9.445  & 400 & VLT/VIMOS & $z$&   20.85\pm0.04\\
9.462  & 60 & VLT/FORS & $z$&   20.69\pm0.11\\
13.440  & 480  & VLT/VIMOS & $z$&  22.30\pm0.28\\
19.463  & 720 & VLT/VIMOS & $z$&  23.37\pm0.48\\
  0.49 & 120 & VISTA/VIRCAM &$Y$ & 17.46\pm0.01\\
  1.47 & 120 & VISTA/VIRCAM &$Y$ & 17.23\pm0.01\\
  2.47 & 120 & VISTA/VIRCAM &$Y$ & 17.51\pm0.02\\
  3.46 & 120 & VISTA/VIRCAM &$Y$ & 17.76\pm0.01\\
  4.46 & 120 & VISTA/VIRCAM &$Y$ & 18.07\pm0.02\\
  6.47 & 120 & VISTA/VIRCAM &$Y$ & 18.71\pm0.04\\
  7.47 & 120 & VISTA/VIRCAM &$Y$ & 19.24\pm0.07\\
  8.46 & 120 & VISTA/VIRCAM &$Y$ & 19.67\pm0.09\\
  9.46 & 120 & VISTA/VIRCAM &$Y$ & 20.09\pm0.14\\
  0.48 & 120 & VISTA/VIRCAM &$J$ & 17.88\pm0.03\\
  0.51 & 120 & VISTA/VIRCAM &$J$ & 17.82\pm0.03\\
  1.46 & 120 & VISTA/VIRCAM &$J$ & 17.45\pm0.01\\
  2.46 & 120 & VISTA/VIRCAM &$J$ & 17.66\pm0.02\\
  3.46 & 120 & VISTA/VIRCAM &$J$ & 17.86\pm0.02\\
  4.46 & 120 & VISTA/VIRCAM &$J$ & 18.08\pm0.03\\
  4.79 & 298 & HST/WFC3-IR  &F110W & 18.26\pm0.01\\
  6.47 & 120 & VISTA/VIRCAM &$J$ & 18.74\pm0.04\\
  7.24 & 298  & HST/WFC3-IR  & F110W& 19.06\pm0.01\\
  7.46 & 120 & VISTA/VIRCAM &$J$ & 19.07\pm0.08\\
  8.45 & 120 & VISTA/VIRCAM &$J$ & 19.69\pm0.09\\
  9.45 & 120 & VISTA/VIRCAM &$J$ & 20.06\pm0.14\\
 10.46 & 120 & VISTA/VIRCAM &$J$ & 20.94\pm0.35\\
 10.55 & 298 & HST/WFC3-IR & F110W& 20.82\pm0.02\\
 11.46 & 120 & VISTA/VIRCAM &$J$ & 21.16\pm0.40\\
4.923 & 298 & HST/WFC3-IR & F160W&  18.063\pm0.03\\
9.427 & 298  & HST/WFC3-IR & F160W&  19.600\pm0.06\\
10.619 & 298  & HST/WFC3-IR & F160W& 20.279 \pm0.09\\
  0.47 & 120 & VISTA/VIRCAM & \Ks & 18.62\pm0.05\\
  0.50 & 120 & VISTA/VIRCAM & \Ks & 18.64\pm0.06\\
  1.32 & 360 & NOT/NOTcam & \Ks & 17.86\pm0.22\\
  1.46 & 120 & VISTA/VIRCAM & \Ks & 17.77\pm0.02\\
  2.45 & 120 & VISTA/VIRCAM & \Ks & 17.67\pm0.03\\
  3.45 & 120 & VISTA/VIRCAM & \Ks & 17.54\pm0.02\\
  4.45 & 120 & VISTA/VIRCAM & \Ks & 17.60\pm0.02\\
  6.46 & 120 & VISTA/VIRCAM & \Ks & 17.84\pm0.03\\
  7.45 & 120 & VISTA/VIRCAM & \Ks & 17.95\pm0.04\\
  8.45 & 120 & VISTA/VIRCAM & \Ks & 18.25\pm0.03\\
  9.45 & 120 & VISTA/VIRCAM & \Ks & 18.49\pm0.05\\
 10.45 & 120 & VISTA/VIRCAM & \Ks & 18.74\pm0.06\\
 12.46 & 120 & VISTA/VIRCAM & \Ks & 19.34\pm0.08\\
 14.46 & 120 & VISTA/VIRCAM & \Ks & 20.02\pm0.13\\
 17.45 & 780  & VLT/HAWK-I & \Ks & 20.77\pm0.13\\
 20.44 & 1140 & VLT/HAWK-I & \Ks & 21.58\pm0.06\\
 21.44 & 1320  & VLT/HAWK-I & \Ks & 21.46\pm0.08\\
 25.44 & 600 & VLT/HAWK-I & \Ks & 22.06\pm0.22\\
\enddata
\tablecomments{Column (1) is the start time of observation with respect to the gravitational wave 
trigger time \citep{GCN21509}. }
\end{deluxetable}

The distance to NGC\,4993 is not well established \citep{hjorth17}.  
The heliocentric velocity is 2930\,km\,s$^{-1}$ \citep[$z\approx0.0098$;][]{levan17}, 
and here we take the distance to be 
$d=40$ Mpc (distance modulus  $\mu=33.01$).   
Thus the peak absolute magnitudes from our
measurements are  $M_{Y,0}=-15.79$ and $M_{K,0}=-15.47$. 

\subsection{Spectroscopy}
\label{sec:spec}
We observed \sss\ with the MUSE integral field spectrograph on the VLT,
which provides optical spectroscopy of both the transient and also the surrounding galaxy (a more detailed description of these data and the analysis of the environment is presented in \citet{levan17}.

Later spectroscopy was obtained with the {\em Hubble Space Telescope}
(\HST) using the Wide-Field Camera 3 Infrared channel (WFC3-IR), with both available grisms, G102  and G141.
These observations 
were pre-reduced by the WFC3 pipeline. The pipeline products were astrometrically calibrated and flat-field corrected, and the diffuse sky background subtracted, using the {\tt python}-based package {\tt grizli}\footnote{\url{https://github.com/gbrammer/grizli}; development in progress}. The significant background contamination, caused by the bright host galaxy, was fitted with a two-dimensional polynomial model in a region around the target spectrum, then subtracted using {\tt astropy} \citep{astropy13}. 
The {\tt grizli} package was then used to optimally extract and combine the spectra from individual exposures. We confirmed these features are robust by comparing the results to extractions from 
the standard {\tt aXe} software.

\begin{figure}[ht!]
\hspace{-2mm}\includegraphics[width=8.7cm,angle=0]{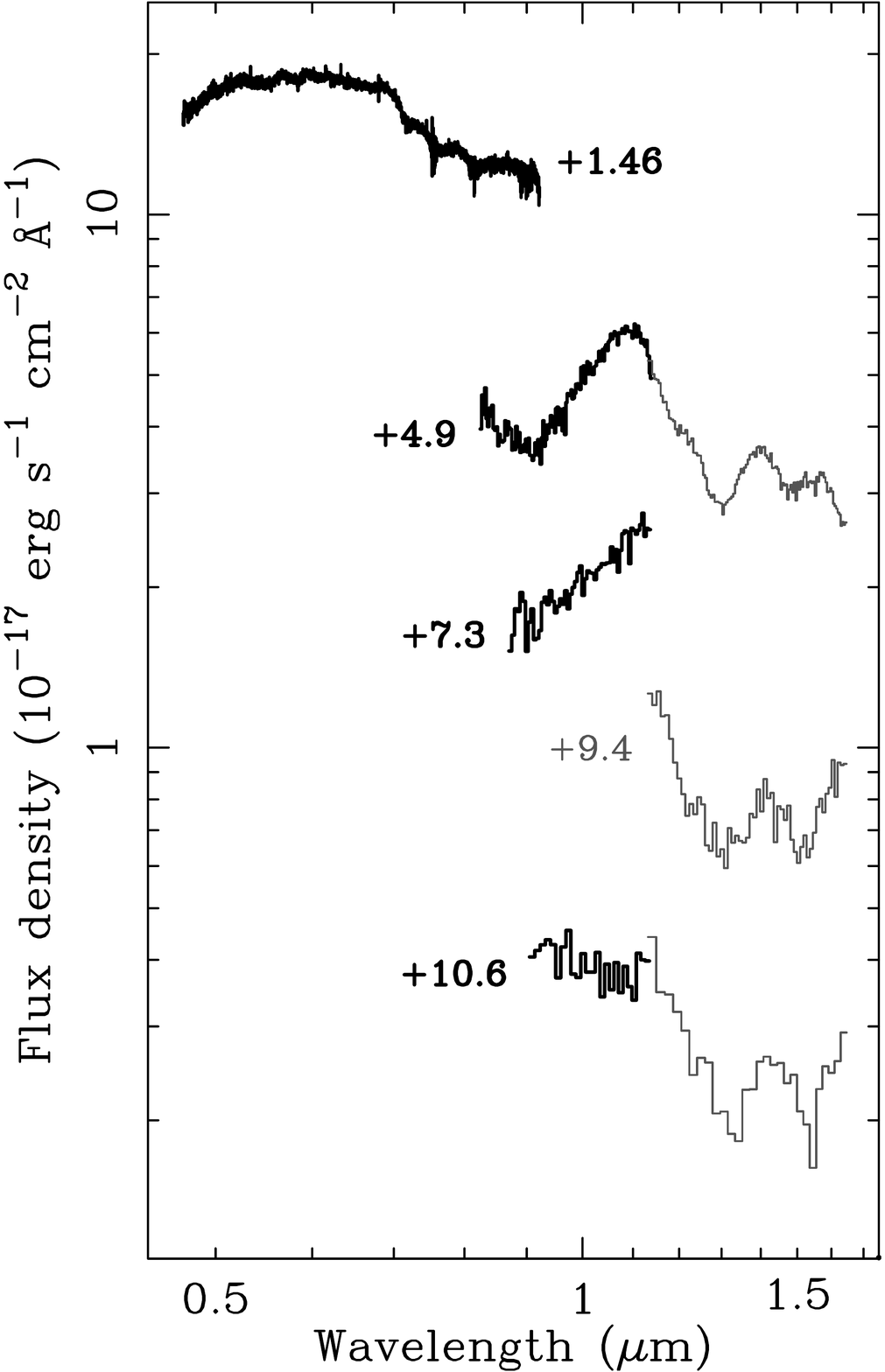}
\caption{VLT/MUSE and \HST\ grism spectra  at five epochs (days post-merger labeled).
The later \HST\ observations have been
rebinned to reduce the noise.
G141 grism spectra are plotted in a lighter line to distinguish them from 
the G102 spectra.
The spectra are scaled to match our photometric observations, but have not been corrected
for Galactic foreground extinction.  Note, since the flux density axis here plots $F_{\lambda}$ the
slopes of the spectra are not directly comparable to Fig.~\ref{fig:sed}
\label{fig:spec}}
\end{figure}

The spectroscopic observations are summarised in Table~\ref{tab:spec}, and the
spectra are plotted in Figure~\ref{fig:spec}.
The first spectrum at roughly 1.5\,d post-merger 
peaks around 0.6\,\micron\ in the optical.
The continuum is smooth, with only weak troughs around 0.55\,\micron, 0.58\,\micron, 0.75\,\micron
and 0.8\,\micron, with a more pronounced break at 0.7\,\micron.
Subsequently, the \HST\ spectra monitor
the behaviour in the near-infrared, and
show that by 5 days the spectrum is dominated by
a prominent peak at $\sim$1.1\,\micron.
Lesser
peaks are apparent at $\sim$1.4\,\micron\ and $\sim$1.6\,\micron, and a weak peak at $\sim$1.22\,\micron.
The breadth of the features is reminiscent of broad-line supernova spectra \citep[e.g.][]{hjorth03}, and their positions, particularly of the $\sim$1.1\,\micron\ peak, matches qualitatively the model spectra of \citet{kasen13} which adopted opacity based on the lanthanide neodymium.
These features appear to be present through the
sequence, although they diminish in significance and move towards
slightly longer wavelengths.
This is consistent with the photosphere moving deeper with time to slower moving ejecta as the faster moving outer layers cool and recombine.
Overall the spectra match well those
seen in the extensive ground-based 
spectroscopic sequence of \citep{pian17},
although the absence of atmospheric absorption, compared to ground-based spectra is particularly beneficial in
revealing clearly the 1.4\,\micron\ feature.

\begin{deluxetable}{cClc}
\tablecaption{Optical and near-IR spectroscopy of  \sss \label{tab:spec}}
\tablecolumns{5}
\tablenum{2}
\tablewidth{0pt}
\tablehead{
\colhead{$\Delta t$ (d)} & \colhead{$t_{\rm exp}$ (s)} & \colhead{Telescope/Camera} & \colhead{Coverage ($\mu$m)}   }
\decimalcolnumbers
\startdata
1.47 & 2600 & VLT/MUSE &  0.48--0.93   \\ %
4.86 & 1812 & \HST/WFC3-IR & 0.8--1.15 (G102)   \\ %
4.93 & 1812 & \HST/WFC3-IR & 1.08--1.7 (G141)    \\ %
7.27 & 1812 & \HST/WFC3-IR & 0.8--1.15 (G102)  \\ %
9.43 & 1812 & \HST/WFC3-IR & 1.08--1.7 (G141)    \\ %
10.52 & 1812 & \HST/WFC3-IR & 0.8--1.15 (G102)   \\ %
10.65 & 1812 & \HST/WFC3-IR & 1.08--1.7 (G141)    \\ %
\enddata
\tablecomments{Column (1) contains start time of observation with respect to gravitational wave 
trigger time. }
\end{deluxetable}

\section{Interpretation}
\label{sec:model}

A natural question is whether any of the light could be due to a synchrotron afterglow, as is generally seen in GRBs.  The absence of early X-ray emission
\citep[for 40\,Mpc distance, $L_{\rm X}<5.24\times10^{40}$\,erg\,s$^{-1}$ at  0.62\,d after the trigger;][]{evans17}, in particular, argues that any afterglow must be faint. A simple extrapolation of the early X-ray limit, assuming conservatively that 
$F_{\nu}\propto \nu^{-1}$, gives $J>19.9$.  This would at most be a minor contribution to the light observed at early times, so we neglect it here.

We currently lack \KN\ model predictions based on a complete set of 
likely elements present, and so conclusions are necessarily
preliminary.  From the large width of the bumps and troughs in
the spectrum, which have roughly $\Delta\lambda/\lambda\sim0.1$ we
may infer characteristic ejecta velocity of up to $v \sim0.1c$, assuming the width is at least partly due to Doppler spreading (see Fig. 4).  
Using this value of the velocity and the light-curve rise time (as well as the decay time of the optical light curves), the ejecta mass $M$ is approximately \citep{arnett80,metzger10}
$$M \sim 5\times 10^{-3}\, M_\odot \left(\frac{0.1\, \textrm{g\, cm}^{-2}}{\kappa} \frac{v}{0.1\, c}
\right),$$
where $\kappa$ is the opacity.  This would suggest that only $10^{50}$ erg of kinetic energy are in the ejecta, despite an energy input of $\sim 10^{53}$ erg during the merger.  

The observed peak isotropic bolometric luminosity of $\sim$few$\times 10^{41}$\,erg\,s$^{-1}$ \citep[integrating between $u$ and \Ks, making use of the UVOT data in ][]{evans17} is much higher than predicted for diffusion through an expanding medium following this initial energy input. Continued powering from radioactive decay is required to explain the observations, and is consistent with the much slower decaying infrared light curve.  Parametrizing the total heating output of radioactive decay as $\epsilon \equiv f M c^2$ \citep[e.g.,][]{metzger10}, we can estimate $f$ as
$$f \sim 10^{-6} \frac{L_\textrm{peak}}{10^{41} \textrm{erg\, s}^{-1}} \frac{0.005 M_\odot}{M}.$$

The fact that the counterpart was bright, even in the UV, in the first $\sim24$\,hr after the merger \citep{evans17},
indicates a high-mass wind with a high $Y_e$ and hence comparatively low opacity ejecta.  This component is likely also dominating the optical
emission at early times.

On the other hand, the relatively rapid decline in the $J$-band compared to the \Ks-band light suggests that the latter must be dominated, at least from a 
few days post-merger, by emission from lanthanide-rich dynamical ejecta, in which nucleosynthesis has proceeded to the third r-process peak.

\subsection{Comparison to theoretical models}

We compare our observations to the two-component models developed in
\cite{wollaeger17}. These models are computed using the multidimensional 
radiative Monte Carlo code 
\SuperNu\,\footnote{\url{https://bitbucket.org/drrossum/supernu/wiki/Home}}
\citep{wollaeger13,wollaeger14,vanrossum16}
with the set of multigroup opacities produced by the Los Alamos suite of 
atomic physics codes \citep{fontes15a,fontes15b,fontes17}. 
Two-component axisymmetric outflow consists of neutron-rich toroidal 
dynamical ejecta \citep{rosswog14}, and a slower spherically-symmetric
homologous outflow with higher electron fraction, broadly referred to as
``wind". 
The $r$-process nucleosynthesis and radioactive heating are computed
using the nuclear network code \WinNet\
\citep{winteler12,korobkin12,thielemann11} with reaction rates compilation
for the finite range droplet model \citep[FRDM,][]{moeller95,rauscher00}.
Coordinate- and time-dependent thermalization of nuclear energy is calculated
using empirical fits developed in \citet{barnes16} and \citet{rosswog17}.

\begin{figure*}[ht!]
\begin{center}
 \includegraphics[width=18.0cm]{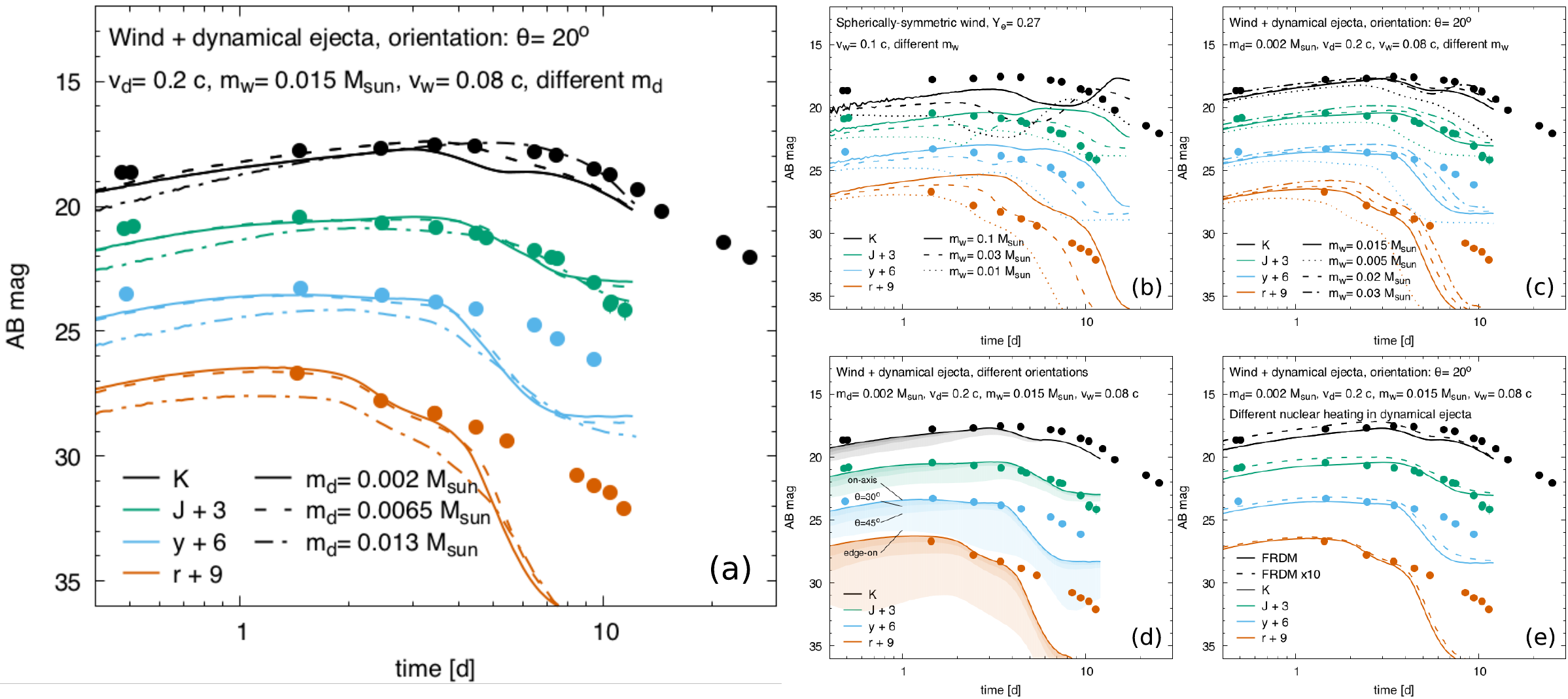}
 \end{center}
 \caption{Effect of varying different parameters of the outflow on the light
 curves in the $ryJK$-bands and spectra: 
 (a) light curves for different dynamical ejecta
 masses with default wind ejecta model;
 (b) light curves for a
 spherically-symmetric wind model with different masses;  (c) light curves for different 
 wind masses; (d) impact of the inclination angle: shaded color bands 
 indicate edge-on, $45^\circ$ and $30^\circ$ inclination, and the continuous
 lines represent on-axis view;  (e) light curves for nuclear heating from FRDM model (default)
 compared to the case with $10\times$ nuclear heating in the dynamical ejecta.
 Filled circles correspond to the observed photometry.}
\label{fig:fits}
\end{figure*}

The models are characterized by five parameters: mass and velocity of the dynamical
ejecta; mass and velocity of the wind outflow; and inclination angle, which characterizes
the remnant orientation. Below we explore a range of these parameters in
comparison with the photometric and spectral observations.
Fig.~\ref{fig:fits} shows the photometry  compared to a few
models with varying individual parameters relative to the baseline model with
dynamical ejecta parameters
$m_{\rm dyn}=0.002\ M_\odot$, $v_{\rm dyn}=0.2\ c$, wind parameters
$m_{\rm wind}=0.015\ M_\odot$, $v_{\rm wind}=0.08\ c$, and orientation angle
$\theta=20^\circ$. Fig.~\ref{fig:fits}b shows observed spectrum compared to 
the synthetic spectrum of the baseline model. We conclude that it provides 
a reasonable fit given the uncertainties in our modeling.

Notice that the wind composition here is moderately neutron-rich, with
initial electron fraction $Y_e=0.27$ \citep[denoted as ``wind~2" in][]{wollaeger17}. Such neutron richness produces a composition of elements
grouped around the first $r$-process peak, and, unlike models with higher
electron fraction \citep[e.g. ``wind~1" in][]{wollaeger17}, supplies sufficient
nuclear heating to explain the observed early emission in the optical bands.
Lanthanides in this composition are synthesized only in trace amounts and do not
have any noticeable impact on the opacity.

Panel (a) in Figure~\ref{fig:fits} compares photometric observations in
the optical $rY$-bands and near-IR $J$\Ks-bands to the light curves 
increasing the dynamical ejecta mass, with other parameters set to their default values. 
Although the fit is not perfect, particularly at later times, the evolution to $\sim$4 days is reasonably well reproduced, and in $J$ and \Ks for longer.
Higher values of the dynamical ejecta mass 
$m_{\rm dyn}=0.013\ M_\odot$ lead to a better fit in the \Ks-band near the peak, 
however the peak epoch shifts to much later time compared to the observed value. 
A higher dynamical ejecta mass also produces dimmer light curves in all bands 
at early times.

Panel (b) shows the spherically symmetric wind-only model with varying mass
$m_{\rm wind}=0.01-0.1\ M_\odot$. The wind-only model qualitatively captures 
the behavior in the $rYJ$-bands, but due to the absence of lanthanides
it underproduces light in the \Ks-band. This demonstrates the need to include
a secondary, neutron-rich outflow with lanthanides, which can redistribute the
emission into the infrared bands.

Panel (c)  shows the impact of adding a small amount
of neutron-rich dynamical ejecta ($m_{\rm dyn}=0.002\ M_\odot$). We can see 
that the infrared bands are reproduced fairly well, however the addition of 
highly opaque component leads to the rapid decay in the $rY$-bands at late 
times when compared to the observations. On the other hand, our models only 
explore limited parameter space in terms of the composition; the late time 
behavior in the optical bands can be cured by tuning the composition of 
neutron-rich component. Since our intent here is only to demonstrate viability 
of the red kilonova hypothesis, adjusting the composition is beyond the scope 
of this paper.

Panel (d) shows the effect of remnant orientation.
Notice that the \Ks-band becomes insensitive to the orientation after
$t=6$~days, indicating that at this epoch the remnant is transparent to the 
infrared emission and the photosphere disappears. Emission in the optical 
bands remains sensitive to the orientation even at $t=10$~days. Nevertheless,
in these conditions, the local thermodynamic equilibrium (LTE) approximation may not be applicable anymore, so 
we stop our simulations beyond this epoch.

Panel (e) shows that higher values of nuclear heating \citep[a possibility pointed out in][]{rosswog17,barnes16} lead to only
marginal increase in brightness due to small mass of the ejecta and
inefficient thermalization in the dilute dynamical ejecta at late times.

\subsection{Comparison to other claimed kilonovae}
\label{sec:comp}

The \KN\ associated with GRB\,130603B was observed at 6.94 days rest-frame post-burst (corresponding to 7.0 days at 40\,Mpc), with an inferred absolute magnitude $M_{\rm J}=-15.35\pm0.2$ ($J=17.66$ at 40\,Mpc).  
This is roughly a factor of three greater than the luminosity in the $J$-band at the equivalent epoch for the kilonova accompanying GW170817/GRB\,170817A,
and could indicate a higher mass of dynamical ejecta, or additional energy injection from the central remnant \citep[cf.][]{kisaka16,gao17}, in that case.

The candidate \KN e discussed by \citet{yang15,jin15} and \citet{jin16} are
more difficult to disentangle from the afterglow contribution, but have absolute AB magnitudes (roughly rest-frame $r$-band) around $-14$ to $-15$ in the range 3--10 days post burst, which is again 
in excess of the emission from \sss.

These comparisons show that some diversity is to be expected, but it bodes well for the detection of dynamically driven emission components in BNS events  at the distances accessible with the advanced GW arrays.

\section{Discussion and Conclusions}
\label{sec:comc}

Our densely sampled optical and near-infrared light curves have revealed
the emergence of a red kilonova following the merger of two neutron stars
in a galaxy at $\sim$40\,Mpc.

Our modeling of the multi-band light curves indicates the presence of at
least two emission components: one with high and one with low opacity.
The former is interpreted as being the ``tidal part" of the dynamical ejecta
that carries the original, very low  electron fraction ($Y_e<0.25$) and results
in ``strong r-process" producing lanthanides/actinides. This conclusion is 
supported by near-IR spectroscopy that shows characteristic features 
expected for high-velocity lanthanide-rich ejecta.
The second component avoids strong r-process via a raised electron 
fraction ($Y_e>0.25$) and may arise from different mechanisms such as 
neutrino-driven winds and/or the unbinding of accretion torus material. 
In either case the ejecta are exposed for much longer to 
high-temperature/high neutrino irradiation conditions which drive them to be more proton-rich.
Taken together, this lends strong observational support to the idea
that compact binary mergers not only produce 
the ``strong r-process" elements, as previously suspected, but
also elements across the entire r-process range.

Although the detection of this event in the Advanced LIGO and Virgo O2 science run is encouraging
for future detection rates, the fact that we have not previously seen a
similar electromagnetic phenomenon in the low redshift universe 
indicates they are rare.
For example, in over 12 years of operation, \swift\ has only located one short-GRB
which could be potentially associated with a host galaxy within
150\,Mpc, and hence might have been comparable to the \sss\ event \citep{levan08}.
In that case no counterpart was found
despite deep optical and near-infrared followup that would have
easily seen a transient as bright as \sss\ unless it were heavily dust obscured.

The arguments for BNS and NS-BH mergers as heavy r-process nucleosynthesis factories \citep{rosswog17,vangioni16}, including from r-process-enriched dwarf galaxies \citep{beniamini16} and the terrestrial abundance of plutonium-244 \citep{hotokezaka15}, are broadly in agreement with other observational constraints: from radio observations of Galactic double neutron-star binaries \citep[e.g.,][]{oshaughnessy10}, from the rate and beaming-angle estimates of short gamma-ray bursts \citep{fong12}, and from population synthesis models of binary evolution \citep[][and references therein]{abadie10}.  

A single observed merger during the Advanced LIGO/Virgo O2 science run is consistent with this rate, and likely also consistent with the absence of previous serendipitous kilonova observations.  On the other hand, the lack of \swift\ observations of other $\gamma$-ray bursts like this one places an upper limit on the rate of similar events.  Future observations will pin down the rate of such events and their typical yields much more precisely, thus establishing their contribution to the heavy-element budget of the universe.

Finally we note that if this system was moderately close to being viewed pole-on 
(e.g. $\lesssim30^{\circ}$), as may be
suggested by the detection of $\gamma$-rays,
more  highly inclined systems could appear fainter in the optical due to the wind component being obscured by 
more widely distributed lanthanide-rich ejecta. If this is the case, then near-infrared observations could
be critical for their discovery.
The depth of our short VISTA observations is such that a similar transient would have been seen straight-forwardly to $\sim3$ times the distance of NGC\,4993, and a more favourable sky location (allowing longer exposures) would have allowed searches to the full BNS
detection range ($\approx200$\,Mpc) 
expected for Advanced LIGO at 
design sensitivity.

\acknowledgments

We thank the staff at ESO, both at Paranal and Garching, for their expert and
enthusiastic support of the observations reported here.

We thank the staff at STScI, in particular, 
Tricia Royle, Alison Vick, Russell Ryan
and Neill Reid for their help in implementing such rapid {\em HST} observations.

The observations with VISTA were gathered by the ESO VINROUGE Survey (198.D-2010). Observations also used data 
from the VISTA Hemisphere Survey (VHS: 179.A-2010).

HST observations were obtained using programs GO 14771 (PI: Tanvir), GO 14804 (PI: Levan), GO 14850 (PI: Troja).

VLT observations were obtained using programmes 099.D-0688, 099.D-0116, 099.D-0622

NRT, KW, PTO, JLO, SR acknowledge support from
STFC.

AJL, DS, JDL acknowledge support from STFC via grant ST/P000495/1. 

NRT \& AJL have project has received funding from the European Research Council (ERC) under the European Union's Horizon 2020 research and innovation programme (grant agreement no 725246, TEDE, Levan).

IM acknowledges partial support from the STFC.

AdUP, CT, ZC, and DAK acknowledge support from the Spanish project AYA 2014-58381-P.  ZC also acknowledges support from the Juan de la Cierva Incorporaci\'on fellowship IJCI-2014-21669, and DAK from Juan de la Cierva Incorporaci\'on fellowship IJCI-2015-26153.

JH was supported by a VILLUM FONDEN Investigator grant (project number 16599).

PDA, SC and AM acknowledge support from the ASI grant I/004/11/3.

SR has been supported by the Swedish Research Council (VR) under grant number 2016- 03657\_3, by the Swedish National Space Board under grant number Dnr. 107/16 and by the research environment grant ``Gravitational Radiation and Electromagnetic Astrophysical Transients (GREAT)" funded by the Swedish Research council (VR) under Dnr 2016-06012. 

PAE acknowledges UKSA support.

The VISTA observations were processed by CGF at the Cambridge Astronomy Survey Unit (CASU), which is funded by the UK Science and Technology Research Council under grant ST/N005805/1.

This research used resources provided by the Los Alamos National
Laboratory Institutional Computing Program, which is supported by the U.S.
Department of Energy National Nuclear Security Administration under
Contract No. DE-AC52-06NA25396.

%

\vspace{5mm}
\facilities{HST(WFC3), VISTA(VIRCAM), VLT(MUSE, HAWK-I, VIMOS, FORS)
}

\end{document}